\documentclass[aps,prstab,twocolumn, groupedaddress,
showpacs,amsmath,amssymb,floatfix,letterpaper]{revtex4}

\usepackage{graphicx}
\usepackage[usenames]{color}

\begin{document} 

\title{Higher Order Spin Resonances in a 2.1~GeV/c Polarized Proton Beam}

\author{M.A.~Leonova}
\author{J.A.~Askari}
\author{K.N. Gordon}
\author{A.D.~Krisch}
\author{J.~Liu}
\author{V.S.~Morozov} 
\altaffiliation[now at ]{Thomas Jefferson National Lab, Newport News, VA 23606}
\author{D.A. Nees}
\author{R.S.~Raymond}
\author{D.W.~Sivers}
\author{V.K.~Wong}
\affiliation{Spin Physics Center, University of Michigan, 
Ann Arbor, MI 48109-1040, USA}

\author{F.~Hinterberger}
\affiliation{Helmholtz-Institut f$\ddot{u}$r Strahlen- und Kernphysik, 
Universit$\ddot{a}$t Bonn, D-53115 Bonn, Germany}


\date{April 25, 2011} 

\begin{abstract} 
Spin resonances can depolarize or spin-flip a polarized beam. 
We studied 1$^{\rm st}$ and higher order spin resonances with stored 2.1~GeV/$c$ 
vertically polarized protons.  The 1$^{\rm st}$ order vertical ($\nu_y$) resonance 
caused almost full spin-flip, while some higher order $\nu_y$ resonances caused partial depolarization. 
The 1$^{\rm st}$ order horizontal ($\nu_x$) resonance caused almost full depolarization, 
while some higher order $\nu_x$ resonances again caused partial depolarization.
Moreover, a 2$^{\rm nd}$ order $\nu_x$ resonance is about as strong as some 3$^{\rm rd}$ order $\nu_x$ 
resonances, while some 3$^{\rm rd}$ order $\nu_y$ resonances are much stronger than 
a 2$^{\rm nd}$ order $\nu_y$ resonance. One thought that $\nu_y$ spin resonances are far stronger than 
$\nu_x$, and that lower order resonances are stronger than higher order; the data do not support this.
\end{abstract} 

\maketitle 
To study the strong interaction's spin dependence with polarized proton beams, 
one must preserve and control the polarization during acceleration 
and storage~\cite{dgcrabb,spin00,spin02,spin04,spin06,spin08}. 
This can be difficult due to many 1$^{\rm st}$ and higher order depolarizing (spin) resonances.  
For vertically polarized beams in flat accelerators, it was thought that vertical 
spin resonances should be stronger than horizontal resonances, and lower order 
resonances should be stronger than higher order resonances~\cite{ssc,fermi}.
There were several theoretical attempts to calculate the strengths of higher order spin 
resonances~\cite{Tepikian,Mane,Hoffstaetter}. 
Some 2$^{\rm nd}$ order and synchrotron-sideband resonances were seen in electron rings~\cite{Johnson} 
and proton rings~\cite{Goodwin}. 
Moreover, a 2$^{\rm nd}$ order proton resonance was studied in detail at IUCF~\cite{Ohmori}. 
We used 2.1~GeV/$c$ polarized protons stored in the COSY synchrotron for a detailed 
experimental study of higher order spin resonances. 
Our preliminary $\nu_y$ data was presented at SPIN 2004~\cite{adk}, but both the $\nu_y$ data and the 
never-presented $\nu_x$ data needed significant reanalysis. 
The properly reanalyzed data presented here suggest that many higher order spin resonances, both 
$\nu_y$ and $\nu_x$, must be overcome to accelerate polarized protons to high energies.

In flat circular rings, a beam proton's spin precesses around the vertical fields of 
the ring's dipole magnets. 
The spin tune $\nu_s = G\:\gamma$ is the number of spin precessions during one turn around the ring, 
where $G = (g - 2)/2$ is the proton's gyromagnetic anomaly and $\gamma$ is its Lorentz energy factor. 
Horizontal magnetic fields can perturb the proton's stable vertical polarization creating a spin 
resonance~\cite{stora,courant,montague,sylee}. 
Spin resonances occur when 
\begin{equation}
\nu_s = k\nu_x + l\nu_y + m,
\vspace{-0.05in}
\end{equation}
where $k$, $l$ and $m$ are integers; 
$\nu_x$ and $\nu_y$ are the horizontal and vertical betatron tunes, respectively. 
Imperfection spin resonances occur when $k~=~l~=~0$. 
Intrinsic spin resonances occur when either $k \neq 0$ or $l \neq 0$, or both; 
the sum $|k|+|l|$ defines each resonance's order. 

The experiment's apparatus, including the COSY storage ring~\cite{maier,lehrach02}, EDDA 
detector~\cite{schwarz,altmeier}, electron cooler~\cite{stein}, 
low energy polarimeter (LEP)~\cite{chiladze}, injector cyclotron, 
and polarized ion source~\cite{eversheim,weidmann,felden}, were shown in Fig.~1 of Ref~\cite{leonova04}. 
The beam from the polarized $H^-$ ion source was accelerated by the cyclotron to 45~MeV 
and then strip-injected into COSY. 

Before this injection, the LEP measured the $H^-$ beam's polarization to monitor its stability. 
The cylindrical EDDA detector~\cite{schwarz,altmeier} measured 
the beam's polarization in COSY after crossing the resonances. 
We reduced its systematic errors by cycling the polarized source~\cite{eversheim,weidmann,felden} 
between the up and down vertical polarization states. 
The measured flat-top polarization, before crossing any resonances, was typically about 75\%.

In the COSY ring, the protons' average circulation frequency $f_c$ was $1.491\:85$~MHz at 2.1~GeV/$c$, 
where their Lorentz energy factor was $\gamma = 2.4514$. 
For these parameters, the spin tune $\nu_s = G\gamma$ was $4.395$. 
During injection, acceleration and at the beginning of the flat-top 
the betatron tunes $\nu_x$ and $\nu_y$ were kept fixed at $3.575$ and $3.525$, respectively. 
This kept both betatron tunes away from any 1$^{\rm st}$, 2$^{\rm nd}$, or 3$^{\rm rd}$ order 
spin resonances on flat-top. 
After reaching the flat-top, we varied the ring quadrupoles' currents to vary either $\nu_y$ 
or $\nu_x$, while keeping the other tune fixed; then we measured the polarization. 

\begin{figure}[t]
\includegraphics[width=0.70\columnwidth]{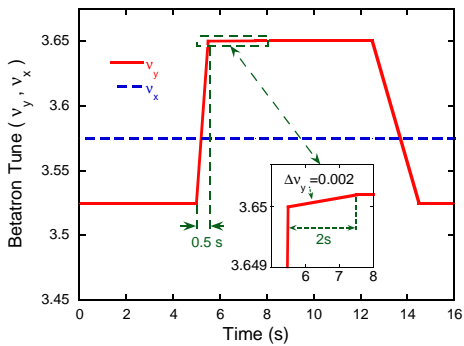}
\caption{Typical $\nu_y$ betatron tune ramps during COSY cycle.
} 
\label{fig:fig1}
\end{figure} 

Figure 1 shows the betatron tunes' behavior in a typical COSY cycle, during the higher order 
vertical ($\nu_y$) spin resonance study; we first ramped $\nu_y$ rapidly from $3.525$ to some 
value between $3.51-3.71$ during 0.5~s. 
Next we slowly ramped $\nu_y$ through a very small tune range of about $0.002$ during 2~s, 
with $\nu_x$ fixed at $3.575$; then we measured the polarization. 
The rapid ramp reduced the effects of the resonances between the injection tune of $3.525$ and 
the start of the slow ramp, while each slow $\nu_y$ ramp enhanced the effect of any spin resonance 
in its very small $\nu_y$ range. 

The Low Energy Polarimeter (LEP) monitored the beam polarization before injection into COSY. 
The measured LEP asymmetries indicated that the initial 
polarization changed during the experiment by about $10\%$. 
Thus, we normalized each final COSY polarization measured 
by EDDA to the measured LEP asymmetry for that data-run. 
The typical duration of each EDDA data-run was 25 min; thus, the LEP data bin sizes were typically 
60~min ($\pm$ 30~min) to include one high count-rate LEP run before each data-run, and one after. 
When needed, the LEP bin size was increased to include a high count-rate LEP run 
both before and after each data run. 

The measured polarizations for the higher order vertical ($\nu_y$) 
spin resonance study are plotted against the final measured $\nu_y$ values in Fig.~2. 
The measured slow betatron tune ramps of about $0.002$ are shown as 
a horizontal bar for each data-point. 
We used Eq.~(1) to calculate the positions of 1$^{\rm st}$, 2$^{\rm nd}$  
and 3$^{\rm rd}$ order resonances that could be studied between 
the half-integer $3.5$ and quarter-integer $3.75$ beam blow-up resonances. 

\begin{figure}[t]
\includegraphics[width=\columnwidth]{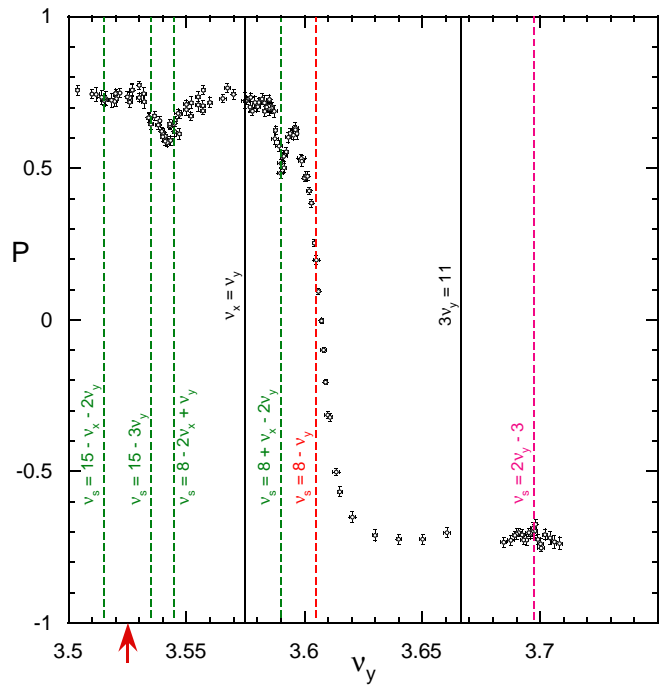}
\caption{Polarization normalized to LEP asymmetry plotted vs. $\nu_y$. 
The measured slow tune ramps of about $0.002$ are shown as horizontal bars. 
The calculated position of each spin resonance is shown by a dashed vertical line and 
the $\nu_x = \nu_y$ and $3\nu_y = 11$ beam-blow-up resonances are shown by solid lines. 
The arrow shows $\nu_y$ at injection.} 
\label{fig:fig2}
\end{figure} 

To test the data's reproducibility, we tried to measure polarizations at the same $\nu_y$ settings several times. 
However, when we precisely measured the $\nu_y$ values after each setting, 
we found that the slow ramps were often not exactly identical at the $\pm 0.0002$ level. 
Thus, Fig.~2 has many partly-overlapping points, which obscure the polarization's behavior near each resonance. 
We tried to clarify Fig.~2 by combining points with nearby $\nu_y$ values, except in the regions 
where the polarization changed very rapidly (between $\nu_y$ values of 3.586 to 3.620). 
We first combined all pairs of points that had $\nu_y$ values within $\mbox{$\delta \nu_y = 0.1 \times 10^{-4}$}$. 
To help ensure that this did not bias the results, we combined the data in both the increasing (Left-to-Right) 
and decreasing (Right-to-Left) $\nu_y$ directions; the two results were identical. 
We then sequentially increased the $\delta \nu_y$ intervals in steps 
of ${0.1 \times 10^{-4}}$; the input data for each step were the output data from the previous step.
The error and position of each newly combined point after each step were the properly weighted 
averages of the errors and positions of the two combined points; each new horizontal bar 
encompassed the slow ramps of both combined points. 

Figure 3 plots polarization vs. $\nu_y$ for the combination interval 
of $\mbox{$\delta \nu_y = 7.6 \times 10^{-4}$}$. 
The 76 combination steps reduced the number of data points from 131 to 95. 
The plot shows clear resonance behavior around several 3$^{\rm rd}$ order resonances, 
but the behavior around the 2$^{\rm nd}$ order resonance is still unclear. 
When we further increased the combination interval size, the polarization's behavior around 
the narrow resonances was broadened excessively, as expected.
 
We observed full spin-flip when the 1$^{\rm st}$ order vertical ($\nu_y$) spin resonance was crossed; 
we also found partial depolarization near 
several 3$^{\rm rd}$ order resonances and possibly near a 2$^{\rm nd}$ order resonance. 
The 3$^{\rm rd}$ order $\nu_s = 8 +\nu_x-2\nu_y$ resonance and the partly overlapping 
3$^{\rm rd}$ order $15 -3\nu_y$ and $8 -2\nu_x+\nu_y$ resonances
appear significantly stronger than the 2$^{\rm nd}$ order $2\nu_y -3$ resonance.
This suggests that many significant 3$^{\rm rd}$ and possibly higher order spin resonances 
must be overcome to accelerate and store polarized protons above 100 GeV.

\begin{figure}[t!]
\includegraphics[width=0.9\columnwidth]{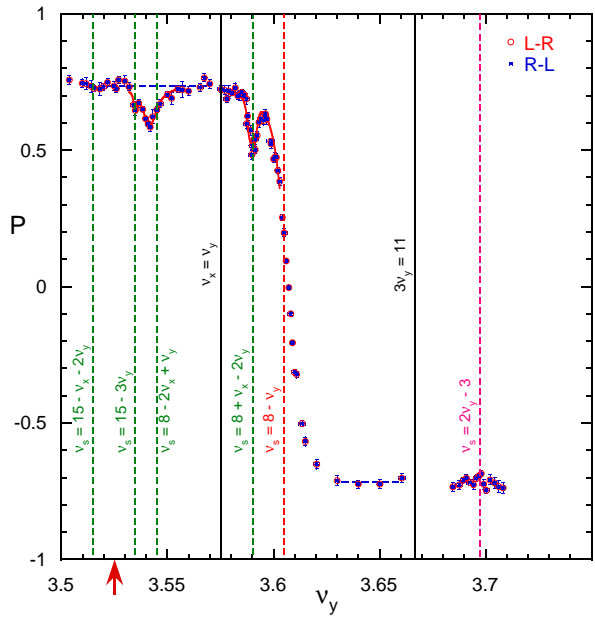}
\caption{Polarization normalized to the LEP asymmetry plotted vs. $\nu_y$. 
(See Fig. 2 caption for more details.) 
The points were combined in steps of ${0.1 \times 10^{-4}}$ up to an interval 
of $\mbox{$\delta \nu_y = 7.6 \times 10^{-4}$}$, except in the $\nu_y$ region of 3.586 to 3.620.
The fits to strong 3$^{\rm rd}$ order spin resonances are shown by the solid red curves. 
The horizontal dashed blue lines show the fits for $P_i$ and $P_f$ for the 1$^{\rm st}$ order $\nu_y$ resonance. 
[See online version with figures expanded by 400-800~\% for details.]} 
\label{fig:fig3} 
\end{figure} 

\begin{figure}[t!]
\includegraphics[width=.9\columnwidth]{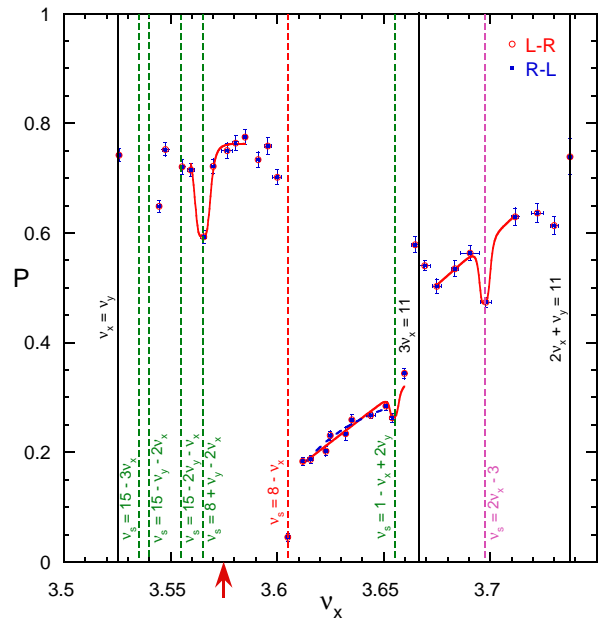}
\caption{Polarization normalized to LEP asymmetry plotted vs. $\nu_x$. 
Only 5 pairs of nearby $\nu_x$ points were combined. 
Dashed vertical lines indicate position of each spin resonance. 
The $\nu_x = \nu_y$, $3\nu_x = 11$ and $2\nu_x + \nu_y = 11$ beam-blow-up 
resonances are shown by solid vertical lines. 
The fits to strong 2$^{\rm nd}$ and 3$^{\rm rd}$ order spin resonances are shown by the solid red curves; 
the dashed blue curve shows the 1$^{\rm st}$ order $\nu_x$ resonance's fit to Eq. (2). 
The arrow shows $\nu_x$ at injection.} 
\label{fig:fig4}
\end{figure} 

We also studied the higher order horizontal ($\nu_x$) spin resonances by using $\nu_x$ 
ramps similar to the $\nu_y$ ramps in Fig. 1, with $\nu_y$ fixed at $3.525$. 
We first rapidly ramped $\nu_x$ from $3.575$ to a value between $3.525-3.74$ in 0.5~s; 
we next slowly ramped $\nu_x$ through a range of about $0.002$ in 2~s; 
then we measured the polarization. 
The rapid ramp again reduced the effects of the resonances between the injection tune of $3.575$ 
and the start of the slow ramp, 
while each slow tune ramp enhanced the effect of the resonance in that small $\nu_x$ range.

The polarizations are plotted in Fig.~4 against $\nu_x$. The 5 pairs of overlapping points were combined, 
as earlier described for Fig.~3.
Figure 4 shows almost full depolarization at the 1$^{\rm st}$ order spin resonance. 
Above this resonance, the polarization increased steadily probably because this fairly strong resonance 
was crossed at increasing $\Delta \nu_x / \Delta t$ rates, which decreased the depolarization~\cite{stora}; 
$\Delta \nu_x / \Delta t$ increased because the ramp time $\Delta t$ was fixed at 0.5 s, while the 
ramp range $\Delta \nu_x$ was increased. 
Thus, we found partial depolarization near a 2$^{\rm nd}$ order $\nu_x$ resonance and near several 
3$^{\rm rd}$ order $\nu_x$ resonances; these $\nu_x$ resonances all seem about equally strong. 
Recall that some 3$^{\rm rd}$ order $\nu_y$ resonances seem significantly stronger than the 
2$^{\rm nd}$ order $\nu_y$ resonance. 

Also note that the polarization increased significantly at the two $\nu_x$ beam-blow-up resonances 
probably because they removed mostly those beam particles with larger betatron amplitudes, 
as supported by the sharp decrease in the precisely measured count rates in EDDA at each blow-up resonance. 
These outside particles were probably more depolarized~\cite{ratner} when crossing the 
strong 1$^{\rm st}$ order $\nu_x$ spin resonance; 
thus, removing them increased the beam's polarization while decreasing its intensity. 

The measured strengths of the 11 resonances, for which we had adequate data, are listed in Table 1. 
We first obtained the very strong 1$^{\rm st}$ order $\nu_y$ resonance's $P_i$ and $P_f$, 
respectively from the left and right horizontal dashed line fits in Fig.~3. 
We then obtained its strength $\varepsilon$ using the measured $P_f/P_i$ and the fast ramp's time 
$\Delta t$ of 0.5~s and $\Delta \nu$ of 0.105 in the Froissart-Stora equation~\cite{stora}:
\begin{equation}
P_f / P_i = 
2 \exp 
\left[ \frac{-(\pi \, \varepsilon)^2 \, f_c}{\Delta \nu/\Delta t}
\right] - 1 
.  
\label{eq:stora}
\end{equation}
We could only set a lower limit on $\varepsilon$ of $240 \times 10^{-6}$ because the 
1$^{\rm st}$ order $\nu_y$ resonance was so strong that the spin was fully flipped 
for our fixed $\Delta t$ of 0.5~s. 
For the strong 1$^{\rm st}$ order $\nu_x$ resonance, the blue dashed curve in Fig.~4 
is the fit of Eq.~(2) to the 8 data points just after crossing it, using $\Delta t$ of 0.5~s 
and $\Delta \nu$ equal to each point's $\Delta \nu_x$ from the $\nu_x$ value at injection. 

For each isolated 2$^{\rm nd}$ and 3$^{\rm rd}$ order resonance, we obtained its dip's depth 
or polarization loss ($P_f/P_i$) by using a $\chi^{\rm 2}$ minimization fit of a 2$^{\rm nd}$ 
order Lorentzian to that resonance's data with a base-line obtained from its nearby points. 
The ($P_f/P_i$) values of the two overlapping 3$^{\rm rd}$ order $\nu_y$ resonances in Fig. 3 
were obtained by a fit using two overlapping Lorentzians and the baseline shown by the 
horizontal dashed blue line. We simultaneously fit the stronger (8 - 2$\nu_x$ + $\nu_y$) 
resonance to a 1$^{\rm st}$ order Lorentzian, with its frequency a variable in the fit, and 
the weaker (15 - 3$\nu_y$) resonance to a 2$^{\rm nd}$ order Lorentzian with its frequency 
held fixed at the calculated value shown by its dashed green line. 
The fits to all 2$^{\rm nd}$ and 3$^{\rm rd}$ order resonances are shown by the solid 
red curves in Figs.~3 and~4. 
Three 2$^{\rm nd}$ and 3$^{\rm rd}$ order resonances had no observable dip at their 
calculated $\nu_x$ or $\nu_y$ value; therefore, the lower limits on their $P_f/P_i$ were 
taken to be 95\%, which was 4 times the average error on straight line fits to the data 
points near these apparently weak resonances. 
We then phenomenologically used ($P_f/P_i$) in Eq. (2) with our fixed experimental 
$\Delta t$ of 2~s and $\Delta \nu$ of 0.002 to obtain 
$\varepsilon$ for each 2$^{\rm nd}$ and $3^{\rm rd}$ order resonance. 

\maketitle 

\begin{table}[h] 
\begin{center}
\caption{Summary of observed resonances' strengths.}
\begin{tabular} {c|c|c|c|c}
\hline \hline
Type & Order & Resonance & $P_f / P_i (\%)$ & $\varepsilon \times 10^{-6}$ \\
\hline
$\nu_y$ & $1^{st}$ & $8-\nu_y$         & $-97.2 \pm 1.4$ & $> 240 $ \\ 
$\nu_y$ & $2^{nd}$ & $2\nu_y-3$        & $ > 95 $        & $< 1.3 $ \\ 
$\nu_y$ & $3^{rd}$ & $15-\nu_x-2\nu_y$ & $ > 95 $        & $< 1.3 $ \\ 
$\nu_y$ & $3^{rd}$ & $15-3\nu_y$       & $92.5 \pm 2.6$  & $ 1.6 \pm 0.4 $ \\ 
$\nu_y$ & $3^{rd}$ & $8-2\nu_x+\nu_y$  & $80.1 \pm 1.7$  & $ 2.7 \pm 0.2 $ \\ 
$\nu_y$ & $3^{rd}$ & $8+\nu_x-2\nu_y$  & $76.9 \pm 4.5$  & $ 2.9 \pm 0.4 $ \\ 
\hline 
$\nu_x$ & $1^{st}$ & $8-\nu_x$         & F-S eq. fit     & $ 41 \pm 2 $ \\ 
$\nu_x$ & $2^{nd}$ & $2\nu_x-3$        & $77.7 \pm 2.2$  & $ 2.8 \pm 0.2 $ \\ 
$\nu_x$ & $3^{rd}$ & $15-3\nu_x$       & \multicolumn{2}{c}{inadequate data} \\
$\nu_x$ & $3^{rd}$ & $15-2\nu_x-\nu_y$ & \multicolumn{2}{c}{inadequate data} \\
$\nu_x$ & $3^{rd}$ & $15-\nu_x-2\nu_y$ & $ > 95 $        & $< 1.3 $ \\
$\nu_x$ & $3^{rd}$ & $8-2\nu_x+\nu_y$  & $77.7 \pm 2.2$  & $ 2.8 \pm 0.2 $ \\ 
$\nu_x$ & $3^{rd}$ & $1-\nu_x+2\nu_y$  & $85.8 \pm 4.7$  & $ 2.2 \pm 0.5 $ \\ 
\hline \hline 
\end{tabular}
\end{center} 
\end{table}
 
There were several theoretical attempts~\cite{Tepikian,Mane,Hoffstaetter} 
to calculate the strengths of higher order spin resonances; 
one~\cite{Tepikian} suggests that odd order resonances may be stronger than even order resonances 
for rings with Siberian snakes. It is not yet clear if these theoretical approaches allow one to 
explain our experimental results. 

In summary, we used 2.1~GeV/$c$ polarized protons stored in the COSY synchrotron 
to study 1$^{\rm st}$ and higher order spin resonances. 
We observed almost full spin-flip when the 1$^{\rm st}$ order 
$8 - \nu_y$ spin resonance was crossed and partial depolarization 
near the 2$^{\rm nd}$ and 3$^{\rm rd}$ order spin resonances. 
We also observed almost full depolarization near the 1$^{\rm st}$ order 
$8 - \nu_x$ spin resonance and partial depolarization near the 
2$^{\rm nd}$ and 3$^{\rm rd}$ order spin resonances. 
It was thought that, for vertically polarized protons in flat accelerators, vertical spin resonances are 
stronger than horizontal resonances, and lower order resonances are stronger than higher order resonances. 
The data suggest that many higher order spin resonances, both horizontal and vertical, must be overcome to 
accelerate polarized protons to high energies; 
these data may help RHIC to better overcome its snake resonances between 100 and 250~GeV/$c$.

We thank COSY's staff for a successful run. 
We thank $\mbox{E.D.~Courant}$, $\mbox{Ya.S.~Derbenev}$, $\mbox{D.~Eversheim}$, 
$\mbox{R.~Gebel}$, $\mbox{A.~Lehrach}$, $\mbox{B.~Lorentz}$, $\mbox{R.~Maier}$, 
$\mbox{Yu.F.~Orlov}$, $\mbox{D.~Prasuhn}$, $\mbox{H.~Rohdje\ss}$, $\mbox{T.~Roser}$, 
$\mbox{H.~Sato}$, $\mbox{A.~Schnase}$, $\mbox{W.~Scobel}$, $\mbox{E.J.~Stephenson}$, 
$\mbox{H.~Stockhorst}$, $\mbox{K.~Ulbrich}$ and $\mbox{K.~Yonehara}$ for help and advice. 
The work was supported by grants from the German BMBF Science Ministry 
and its JCHP-FFE program at COSY.

\begin{figure*}[t!]
\includegraphics[scale=0.9, angle=270]{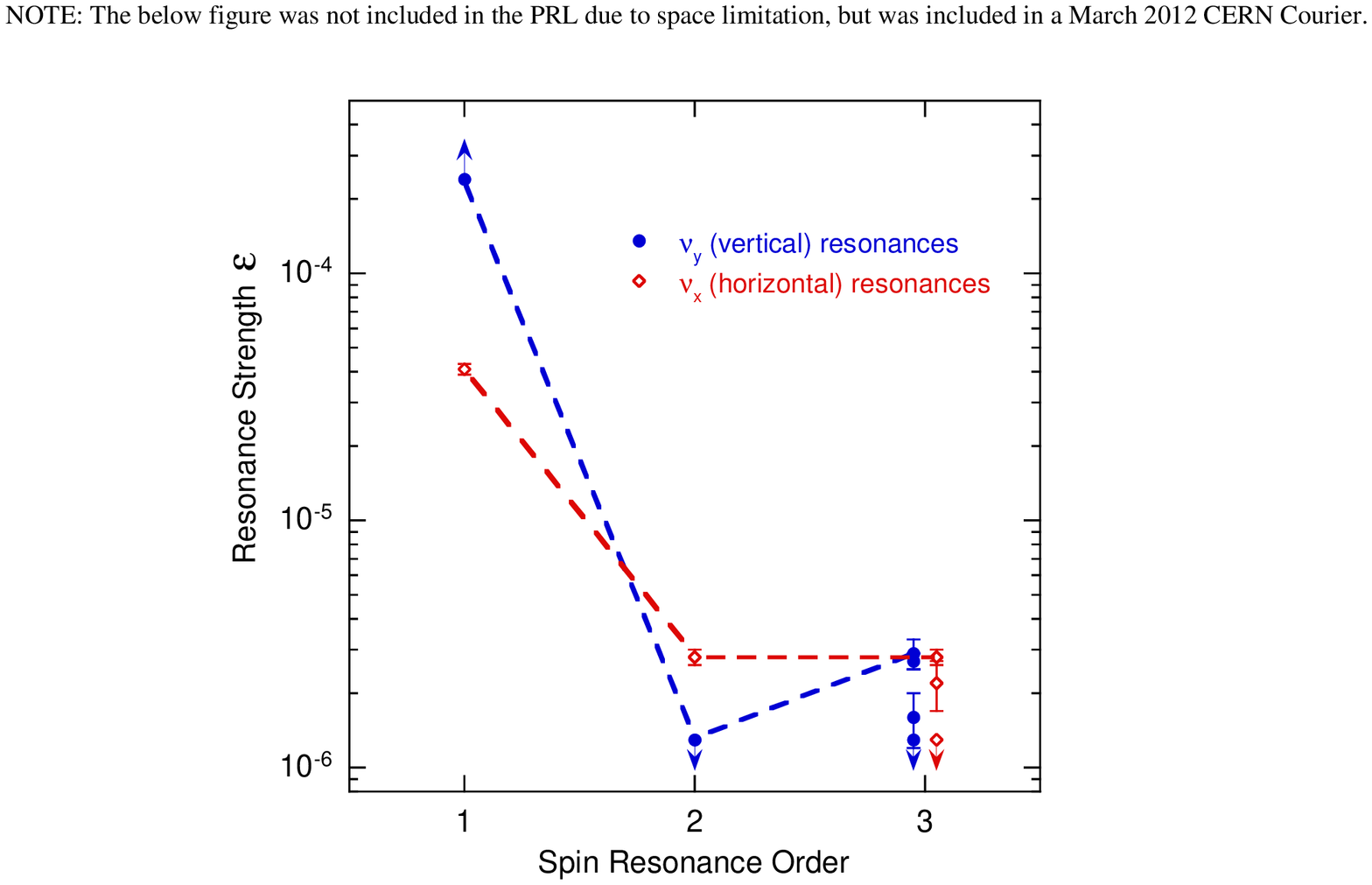}
\protect{\caption{The measured spin resonance strengths obtained from Fig. 3 and Fig. 4 are plotted against the resonance order. The up arrow indicates a lower limit on the resonance strength, while the down arrows each indicate an upper limit. The paper was submitted to PRL on 25 April 2011 and published on 13 February 2012 PRL 108, 074801 (2012) without Fig. 5 due to space limitation.}}
\label{fig:fig5}
\end{figure*} 

\end{document}